# MICROREONATOR-STABILIZED EXTENDED CAVITY DIODE LASER FOR SUPERCAVITY FREQUENCY STABILIZATION

JINKANG LIM,[1,*] ANATOLIY A. SAVCHENKOV,[2] ANDREY B. MATSKO,[2] SHU-WEI HUANG,[1] LUTE MALEKI,[2] AND CHEE WEI WONG[2]

[1]*Mesoscopic Optics and Quantum Electronics Laboratory, University of California, Los Angeles, CA 90095, USA*
[2]*OEwaves Inc., 465 North Halstead Street, Suite 140, Pasadena, CA 91107, USA*
*Corresponding author: jklim001@ucla.edu

**We demonstrate a simple, compact, and cost-effective laser noise reduction method for stabilizing an extended cavity diode laser to a 3×10$^5$ finesse mirror Fabry-Pérot (F-P) cavity corresponding to resonance linewidth of 10 kHz using a crystalline MgF$_2$ whispering gallery mode microresonator (WGMR). The laser linewidth is reduced to sub-kHz such that a stable Pound-Drever-Hall (PDH) error signal is built up. The wavelength of the pre-stabilized laser is tunable within a large bandwidth covering the high reflection mirror coating of a F-P supercavity. © 2016 Optical Society of America**

http://dx.doi.org/10.1364/OL.99.099999

Ultrastable coherent lasers are indispensable for optical frequency standards and fundamental physics research such as high-resolution spectroscopy and optical atomic clock as well as tests of fundamental physical constants [1-4]. High finesse optical F-P mirror cavities are used for the frequency stabilization and noise-reduction of laser sources. Such cavities have finesses higher than 10$^5$ and the corresponding quality factors ($Q$) of greater than 10$^{10}$. A laser frequency is then stabilized to the F-P supercavity via Pound-Drever-Hall (PDH) technique [5] that provides a high signal-to-noise ratio. However, the spectral linewidth of conventional diode lasers is much wider than the supercavity resonance bandwidth, which leads to severe distortions or oscillations in the PDH error signal implying that the direct laser locking is not trivial and requires a high bandwidth feedback servo.

The dynamic response of ultrahigh finesse F-P cavities have been investigated to achieve low-noise laser frequency stabilization [6-8], which has shown that the distortion of a PDH error signal is closely related to the frequency noise of the laser and the cavity resonance bandwidth at the given ring-down time. This indicates that the laser frequency noise should be reduced before the F-P supercavity and thus the laser linewidth needs to be similar or less than the bandwidth of the F-P supercavity mode. This has normally been accomplished by utilizing a pre-stabilization cavity possessing a lower $Q$-factor than that of the F-P supercavity. For instance, F-P mirror bulk cavities are conventionally used for this purpose [9,10] but they need an alignment and proper mode matching optics. WGMRs have been utilized for narrowing the linewidth of diode lasers using optical feedback via Rayleigh backscattering [11,12] or optical filtering in the laser cavity [13] with compact platforms. In this paper, we propose and demonstrate a crystalline MgF$_2$ high-$Q$ WGMR as a pre-stabilization cavity by actively stabilizing a laser frequency to it [14]. A properly designed WGMR not only reduces the linewidth of a laser by orders of magnitude but also mitigates the laser frequency drift by more than an order of magnitude. The WGMR is mechanically robust and the broad transmission window (ultraviolet to mid-infrared) such that it is appropriate to apply for multiple laser wavelength stabilization with a single device. This promises more compact, alignment-free, mode-matching-optics-free, and cost-effective pre-stabilization cavities with high fidelity and performance.

The F-P supercavity mirror has relatively broad high refection coating bandwidth, for example Fig. 1(b), but the commercial F-P cavity stabilized lasers cover only tiny part of its coating bandwidth (< 10 GHz) because of the mode-hop-free tunable range of the built-in narrow-linewidth external cavity diode laser. However, applications for measuring phase noise of lasers or for interrogating atomic resonances need a change of the stable laser frequency to achieve a detectable radio frequency (rf) beatnote signal and a strong resonance signal respectively. Here we show that this can be done by utilizing a largely tunable extended-cavity diode laser (ECDL) stabilized to the high-$Q$ WGMR as an external laser source. Figure 1(a) illustrates our experimental set-up. A tunable ECDL (new focus TLB-6700) possessing < 1 MHz linewidth at 100 ms integration time is stabilized to a WGMR [15,16]. The WGMR is made of crystalline $MgF_2$, which has the free spectral range (FSR) of 10.7 GHz and $Q$ of $10^9$ corresponding to ~100 kHz resonance bandwidth. The temperature of the WGMR is read out by a 10 kΩ thermistor sensor and controlled by a P-I feedback via a Peltier-type thermo-electric cooler (TEC) attached to the WGMR, which maintains the cavity temperature below 10 mK in hours. The laser light is delivered by a single mode polarization maintaining fiber with a micro-lens collimator and coupled into the WGMR via a glass prism with ~50 % efficiency. To prevent the back-reflection, a micro-isolator is also installed. The transmitted light through the WGMR is detected by a built-in photodetector at the output. The whole pre-stabilization unit is packaged in an aluminum box with a compact footprint (volume of 24 $cm^3$).

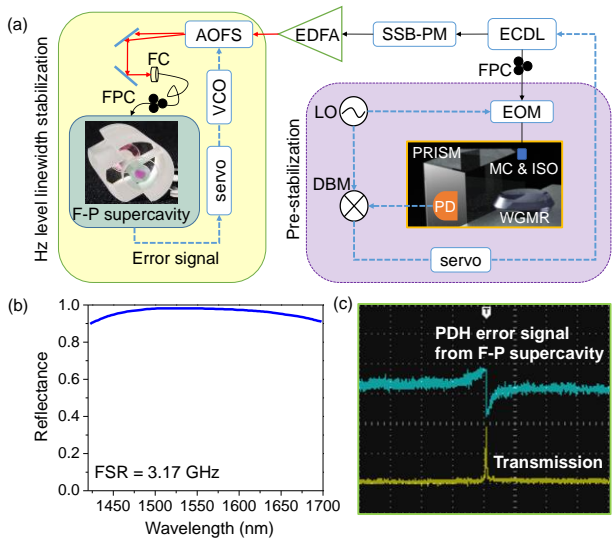

**Fig. 1.** (a) The ECDL is stabilized to the WGMR via PDH locking. The PDH error signal is created by measuring the transmitted light and phase sensitive demodulation at the DBM. When the laser frequency is close to the resonance, the transmitted power decreases because the light is stored in the WGMR. The pre-stabilized ECDL is frequency shifted by a SSB-PM reserved for matching mode between two different FSR cavities: WGMR (10.7 GHz) and F-P cavity (3.17 GHz). The pre-stabilized ECDL is amplified and launched into the super-cavity. The PDH error signal is used to control the applying frequency to the AOFS via a VCO. (b) The F-P cavity mirror reflectance vs. wavelength. Ultrahigh reflection bandwidth 1545 nm ± 1.5 % (c) The triggered error signal and transmission signal from the F-P cavity. FPC: fiber-optic polarization controller, EOM: electro-optic modulator, MC: micro-collimator, ISO: isolator, DBM: double balanced mixer, LO: local oscillator, PD: photodetector, FC: fiber collimator, AOFS: acousto-optic frequency shifter, VCO: voltage-controlled oscillator.

The ECDL is stabilized to the WGMR via PDH lock technique. The light is phase-modulated via an electro-optic modulator with a 10 MHz rf sinusoidal signal and coupled into a 90/10 beam splitter. The light at 90 % arm is sent to the WGMR and 10 % arm is used to monitor the input power into the WGMR. The optical power of 50 μW is coupled into the WGMR and we find its resonance by slowly changing the ECDL current with a ramping frequency modulation. The transmitted light through the WGMR is recorded in a photodetector and the phase sensitive demodulation at a double balanced mixer creates a PDH error signal. The error signal is fed to the feedback servo (new focus, LB1005) and the feedback signal from the servo controls the laser current with the feedback bandwidth of ~1 MHz.

Figure 2 shows the frequency noise power spectral density (FNPSD) and Allan deviation (AD) of the ECDL after the pre-stabilization. These characteristics are measured by heterodyne-beating the pre-stabilized ECDL against a reference laser possessing 1 Hz linewidth and < 0.1 Hz/s drift rate (Stable Laser Systems). We measure the FNPSDs of the pre-stabilized ECDL at three different wavelengths as shown Fig 2(a). For the three wavelengths, frequency noises are fairly same near the carrier and above 1 kHz offset frequencies. The peaks stem from the harmonics of the 60 Hz electrical power-line noise and acoustic/vibrational noise. Some differences are observed at the offset frequency ranging from 30 Hz to 1 kHz, which is likely attributed to different frequency noise characteristics of the ECDL at the wavelength regimes. The linewidth of the pre-stabilized ECDL is deduced from the FNPSD by calculating effective linewidth via the $\beta$-separation line method [17]. This approach implements a simple geometric line (dashed gray line in Fig 2(a)) based on the low-pass filtered white noise to determine the laser linewidth for an arbitrary noise spectrum. The FNPSD can be separated into two regions that affect the line shape. In the region above the $\beta$-separation line, the noise contributes to the central part of the line shape. By integrating the FNPSD up to the offset frequency, which is the region above the $\beta$-separation line, we determine that a sub-kHz linewidth is possible when the 60 Hz harmonic noise spikes are not considered as shown in Fig 2(a) inset. In addition to the frequency noise reduction, the frequency instability is also improved. The AD of the pre-stabilized ECDL illustrated by Fig 2(b) shows the stability is enhanced by more than an order of magnitude compared with that of the free-running ECDL. The measured AD is 1.9 (8.3) kHz at 0.1 (1) s averaging time. Although the measurement shows a long-term frequency drift and associated low frequency noise, this drift and noise can be compensated by locking the pre-stabilized ECDL to an F-P supercavity.

The pre-stabilized ECDL is launched to a single-side-band phase modulator (SSB-PM) that is reserved for tuning the laser frequency to a resonance mode of the F-P supercavity. We used the SSB-PM for convenience to control the laser frequency but in principle the SSB-PM can be eliminated and the cavity resonance modes can be matched by using the temperature control of the WGMR. The resonance mode of the WGMR can be tuned by changing the cavity temperature (350 MHz/°C). The light is then deflected by an acousto-optic frequency shifter (AOFS) with 80 % efficiency and coupled into a single mode fiber and delivered into the supercavity (Stable Laser Systems) placed in an ion-pumped high vacuum ($10^{-8}$ Torr) chamber with intense environmental noise shielding. The frequency bandwidth of the cavity resonance is approximately 10 kHz. This F-P cavity mirror has the ultrahigh reflection coating from 1525 nm 1575 nm (Fig. 1(b)) and the F-P cavity sat on an ultralow-expansion spacer allows for ultra-stable laser frequency stabilization supporting a 1 Hz linewidth and 0.1 Hz/s drift laser operating at C- and L-optical bands.

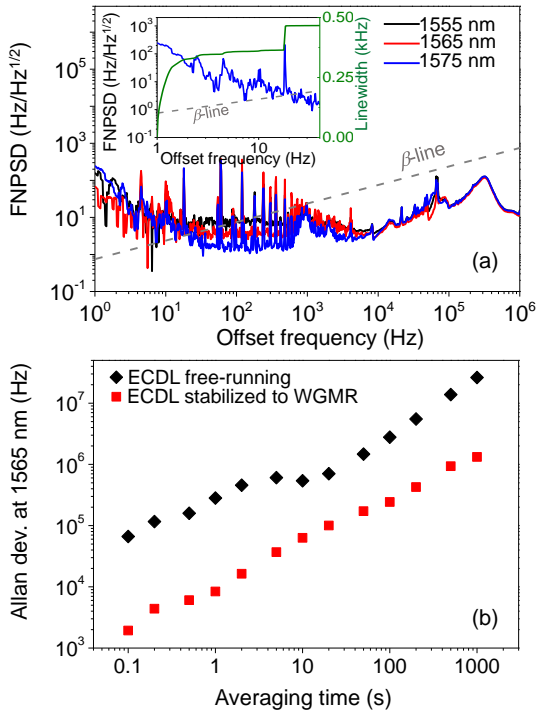

**Fig. 2.** (a) FNPSD of the pre-stabilized ECDL at three different wavelengths. Dashed gray is the $\beta$-separation line. The spikes stem from the 60 Hz harmonics of the electric power-line and acoustic and vibration noise from the environment. Inset: the calculated linewidth of the pre-stabilized ECDL at 1575 nm. (b) The Allan deviation of the WGMR pre-stabilized ECDL (red) at 1565 nm. Frequency stability is improved by more than an order of magnitude compared to that of free-running ECDL (black).

To lock the pre-stabilized ECDL to a high finesse F-P cavity mode, another PDH loop is applied. A cavity resonance of the F-P cavity is interrogated by tuning the laser frequency with the SSB-PM or with the AOFS and the created error signal is used to control the AOFS via a voltage-controlled oscillator (VCO). Due to the reduced frequency drift of the pre-stabilized ECDL, the PDH error signal and the transmission signal can be readily triggered by applying a slow ramping voltage signal to the VCO driving the AOFS as illustrated in Fig. 1(c). The pre-stabilized ECDL frequency is eventually controlled by the AOFS and stabilized to the F-P supercavity.

Since the laser frequency noise after this F-P super-cavity stabilization is better than any other laser in our lab, we first characterize the laser noise by measuring the phase noise of a frequency comb referenced to this F-P supercavity stabilized laser. One of the comb tooth of the 250 MHz fiber comb (Menlosystems) is tightly stabilized to the super-cavity stabilized laser as an optical reference by controlling the comb repetition frequency ($f_{rep}$) with ~300 kHz feedback bandwidth and the carrier envelope offset (CEO) frequency is also stabilized to a low noise rf source. The single side band (SSB) phase noise of the frequency divided signal at 1.25 GHz is measured (olive) as illustrated in Fig. 3. The phase noise of the fully stabilized comb at 5$^{th}$ $f_{rep}$ harmonic is limited by the signal source analyzer (Agilent E5052A) noise floor (red squares). For comparison, we also measure the phase noise of the comb stabilized to the F-P cavity-stabilized laser with the 10 kHz linewidth laser diode (RIO) together (cyan). The measurement of offset frequency noise higher than 100 kHz is limited by the photodetector shot noise. (New focus, 1611FC-AC).

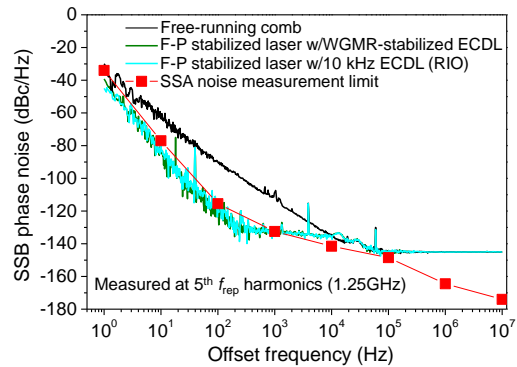

**Fig. 3.** SSB phase noise of the 5$^{th}$ $f_{rep}$ harmonics (1.25 GHz) of the fully stabilized fiber frequency comb when the comb is free-running (black), referenced to F-P supercavity stabilized laser with WGMR pre-stabilized laser (olive), and one with the 10 kHz ECDL (cyan). The red squares are the noise measurement limit of the signal source analyzer (SSA) for a 1.25 GHz carrier frequency.

To avoid the signal source analyzer limitation, we simultaneously stabilize a 3 kHz linewidth self-injection locked continuous wave (cw) laser (OEwaves Inc.) and the fiber comb using a filtered comb tooth via a 1200 grooves/mm grating to the F-P supercavity stabilized laser with enough feedback bandwidths respectively. We, then, beat the cw laser against a comb tooth (out-of-loop beat) as illustrated in Fig. 4(a). The measured relative linewidth of the beatnote recorded by an rf spectrum analyzer shows 1 Hz and the measurement accuracy is limited by the resolution bandwidth of the spectrum analyzer (Fig. 4(b) inset). The AD of this beat signal is also measured. Since both the fiber comb and the cw laser are stabilized to a common optical reference, they ideally have common mode noise. Indeed, the measurement shows the sub-Hz AD over 1 s averaging time as shown in Fig 4(b). Furthermore, we compare the phase noise between the 1 Hz F-P cavity laser with a 10 kHz linewidth ECDL (RIO), used for the commercial 1 Hz F-P cavity laser, and the F-P cavity-stabilized laser with our WGMR pre-stabilized tunable ECDL. Both use the same F-P supercavity as a reference. By locking the cw laser and the fiber comb to the two different 1 Hz F-P reference lasers and measuring their out-of-loop beatnote, the phase noise of the two different F-P cavity stabilized lasers are evaluated respectively shown in Fig. 4(c). They are almost identical, which implies that the noise of both lasers is limited by the noise of the F-P reference cavity. A bump around ~300 kHz offset frequency originates from the fiber comb locking electronics. Note that above 1 MHz offset frequency, the F-P cavity-stabilized laser with the WGMR-stabilized ECDL shows less noise probably owing to the noise filtering by the WGMR.

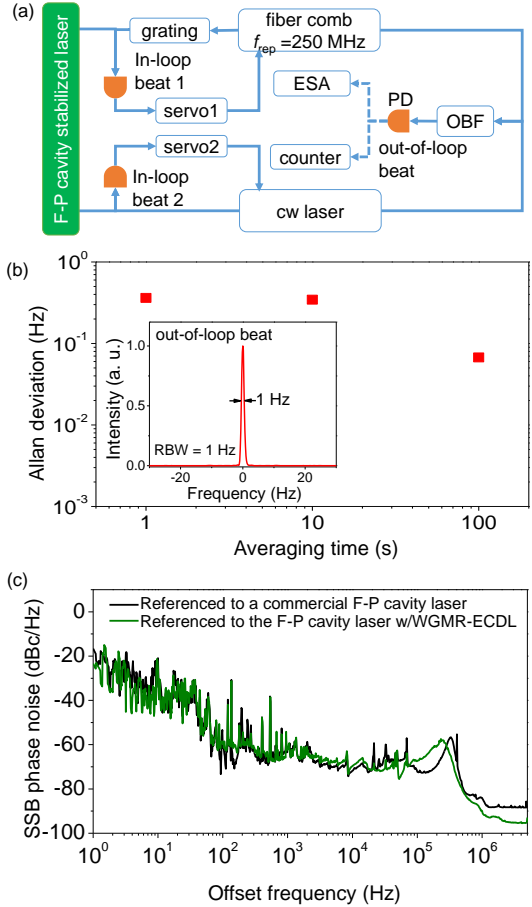

**Fig. 4.** (a) One of the fiber comb tooth is referenced to the F-P cavity stabilized laser and the CEO frequency of the fiber comb is also stabilized to a microwave oscillator. A cw laser is simultaneously referenced to the F-P cavity stabilized laser. The out-of-loop beatnote between one of the fiber comb teeth and the cw laser are detected for the relative linewidth measurement. OBF: optical bandpass filter, ESA: electric spectrum analyzer. (b) Allan deviation of the beatnote. Inset: the instrument-limited linewidth of the beatnote at the 1 Hz resolution bandwidth (RBW). (c) SSB phase noise of the beatnote between one of the fiber comb teeth and the cw laser, both referenced to the F-P cavity stabilized laser with the WGMR-stabilized ECDL (olive) and the commercial F-P cavity stabilized laser (black) using the 10 kHz ECDL (RIO) respectively.

In summary, we propose and demonstrate a broadly tunable pre-stabilization WGMR allowing the sub-kHz linewidth and AD of 1.9 (8.3) kHz at 0.1 (1) s averaging time for the FP supercavity laser frequency stabilization. A stable error signal is built up within the cavity storage time and thus the ECDL is faithfully stabilized to the F-P cavity leading to the cavity noise limited performance. The phase noise of the super-cavity stabilized laser is first characterized with the frequency divided signal via an optical frequency comb referenced to this laser by measuring the 5$^{th}$ $f_{rep}$ harmonic, whose phase noise measurement is limited by the signal source analyzer. The linewidth and phase noise are further characterized by measuring the out-of-loop beatnote between a cw laser and a filtered comb tooth, where both cw laser and comb are stabilized to the F-P supercavity stabilized ECDL assisted by the WGMR. We measured the resolution-limited relative linewidth of 1 Hz. The phase noise is -20 dBc/Hz at 1 Hz, -40 dBc/Hz at 10 Hz, and -65 dBc/Hz at 100 Hz for 191 THz carrier and noise filtering of the pre-stabilization WGMR helps reducing the phase noise above 1 MHz offset frequency. The MgF$_2$ crystal has a broad transmission window, therefore in principle, this WGMR could be used to reduce laser noise in other wavelengths ranging from UV to mid-infrared. The WGMR-assisted pre-stabilization promises more compact, alignment-free, mode-matching-optics-free, and cost-effective method to achieve an F-P supercavity laser stabilization.

**Funding.** Air Force Research Laboratory (FA9453-14-M-0090), the Air Force Young Investigator award (FA9550-15-1-0081 to S.W.H.), and the Office of Naval Research (N00014-14-1-0041)

**Acknowledgement.** We thank discussions with Abhinav K. Vinod.

**References**

1. B. J. Bloom, T. L. Nicholson, J. R. Williams, S. L. Campbell, M. Bishof, X. Zhang, S. L. Bromley, and J. Ye, Nature **506**, 71-75 (2014)
2. A. D. Ludlow, T. Zelevinsky, G. K. Campbell, S. Blatt, M.M. Boyd, M. H. G. Mirandd, M. J. Martin, J. W. Thomsen, S. M. Foreman, Jun Ye, T. M. Fortier, J. E. Stalnaker, S. A. Diddams, Y. Le Coq, Z. W Barber, N. Poli, N. D. Lemke, K. M. Beck, and C. W. Oates, Science **319**, 1805-1808 (2008)
3. R. M. Godun, P. B. R. Nisbet-Jones, J. M. Jones, S. A. King, L. A. M. Johnson, H. S. Margolis, K. Szymaniec, S. N. Lea, K. Bongs, and P. Gill, Phys. Rev. Lett. **113**, 210801 (2014)
4. The LIGO Scientific collaboration, Rep. Prog. Phys. **72**, 076901 (2009)
5. R. W. P. Drever, J. L. Hall, F. V. Kowalski, J, Hough, G. M. Ford, A. J. Munley, and H. Ward, Appl. Phys. B **31**, 97 (1983)
6. M. J. Lawrence, B. Willke, M. E. Husman, E. K. Gustafson, and R. L. Byer, J. Opt. Soc. Am. B. **16**, 523-532 (1999)
7. A. Schoof, J. Grunert, S. Ritter, and A. Hemmerich, Opt. Lett. **26**, 1562-1564 (2001)
8. H. Rohde, J. Eschner, F. Schmidt-Kaler, and R. Blatt, J. Opt. Soc. Am. B. **19**, 1425-1429 (2002)
9. B. C. Yoyngm F. C. Cruz, W. M. Itano, and J. C. Bergquist, Phys. Rev. Lett. **82**, 3799-3802 (1999)
10. A. D. Ludlow, X. Huang, M. Notcutt, T. Zanon-Willette, S. M. Foreman, M. M. Boyd, S. Blatt, and J. Ye, Opt. Lett. **32**, 641-643 (2007)
11. V. V. Vassiliev, V. L. Velichansky, V. S. Ilchenko, M. L. Gorodetsky, L. Hollberg, and A. V. Yarovitsky, Opt, Commun. **158**, 305-312 (1998)
12. W. Liang, V. S. Ilchenko, A. A. Savchenkov, A. B. Matsko, D. Seidel, and L. Maleki, Opt. Lett. **35**, 2822-2824 (2010)
13. B. Sprenger, H. G. L. Schwefel, Z. H. Lu, S. Svitlov, and L. J. Wang, Opt. Lett. **35**, 2870-2872 (2010)
14. J. Lim, A. A. Savchenkov, E. Dale, W. Liang, D. Eliyahu, V. Ilchenko, A. B. Matsko, L. Maleki, and C. W. Wong, arxiv.org/abs/1701.05285
15. I. S. Grudinin, A. B. Matsko, A. A. Savchenkov, D. Strekalov, V. S. Ilchenko, and L. Maleki, Opt. Commun. **265**, 33-38 (2006)
16. A. A. Savchenkov, A. B. Matsko, V. S. Ilchenko, N. Yu, and L. Maleki, *J. Opt. Soc. Am. B*. **24**, 2988-2997 (2007)
17. G. D. Domenico, S. Schilt, and P. Thomann, App. Opt. **49**, 4801-4807 (2010)

## References

1. B. J. Bloom, T. L. Nicholson, J. R. Williams, S. L. Campbell, M. Bishof, X. Zhang, S. L. Bromley, and J. Ye, "An optical lattice clock with accuracy and stability at the 10-18 level," Nature **506**, 71-75 (2014)
2. A. D. Ludlow, T. Zelevinsky, G. K. Campbell, S. Blatt, M.M. Boyd, M. H. G. Mirandd, M. J. Martin, J. W. Thomsen, S. M. Foreman, Jun Ye, T. M. Fortier, J. E. Stalnaker, S. A. Diddams, Y. Le Coq, Z. W Barber, N. Poli, N. D. Lemke, K. M. Beck, and C. W. Oates, "Sr Lattice clock at $1 \times 10^{-16}$ fractional uncertainty by remote optical evaluation with a Ca clock," Science **319**, 1805-1808 (2008)
3. R. M. Godun, P. B. R. Nisbet-Jones, J. M. Jones, S. A. King, L. A. M. Johnson, H. S. Margolis, K. Szymaniec, S. N. Lea, K. Bongs, and P. Gill, "Frequency ratio of two optical clock transition in $^{171}Yb^+$ and constraints on the time variation of fundamental constants," Phys. Rev. Lett. **113**, 210801 (2014)
4. The LIGO Scientific collaboration, "LIGO: the laser interferometer Gravitational-wave observatory," Rep. Prog. Phys. **72**, 076901 (2009)
5. R. W. P. Drever, J. L. Hall, F. V. Kowalski, J, Hough, G. M. Ford, A. J. Munley, and H. Ward, "Laser phase and frequency stabilization using an optical resonator," Appl. Phys. B **31**, 97 (1983)
6. M. J. Lawrence, B. Willke, M. E. Husman, E. K. Gustafson, and R. L. Byer, "Dynamic response of a Fabry-Perot interferometer," J. Opt. Soc. Am. B. **16**, 523-532 ( 1999)
7. A. Schoof, J. Grunert, S. Ritter, and A. Hemmerich, "Reducing the linewidth of a diode laser below 30 Hz by stabilization to a reference cavity with a finesse above $10^5$," Opt. Lett. **26**, 1562-1564 (2001)
8. H. Rohde, J. Eschner, F. Schmidt-Kaler, and R. Blatt, "Optical decay from a Fabry-Perot cavity faster than the decay time," J. Opt. Soc. Am. B. **19**, 1425-1429 (2002)
9. B. C. Yoyngm F. C. Cruz, W. M. Itano, and J. C. Bergquist, "Visible lasers with Subhertz linewidths," Phys. Rev. Lett, **82**, 3799-3802 (1999)
10. A. D. Ludlow, X. Huang, M. Notcutt, T. Zanon-Willette, S. M. Foreman, M. M. Boyd, S. Blatt, and J. Ye, "Compact, thermal-noise-limited optical cavity for diode laser stabilization at $1 \times 10^{-15}$," Opt. Lett. **32**, 641-643 (2007)
11. V. V. Vassiliev, V. L. Velichansky, V. S. Ilchenko, M. L. Gorodetsky, L. Hollberg, and A. V. Yarovitsky, " Narrow-line-width diode laser with a high-*Q* microsphere resonator," Opt, Commun. **158**, 305-312 (1998)
12. W. Liang, V. S. Ilchenko, A. A. Savchenkov, A. B. Matsko, D. Seidel, and L. Maleki, "Whispering-gallery-mode-resonator-based ultranarrow linewidth external-cavity semiconductor laser," Opt. Lett. **35**, 2822-2824 (2010)
13. B. Sprenger, H. G. L. Schwefel, Z. H. Lu, S. Svitlov, and L. J. Wang, "$CaF_2$ whispering-gallery-mode-resonator stabilized-narrow-linewidth laser," Opt. Lett. **35**, 2870-2872 (2010)
14. J. Lim, A. A. Savchenkov, E. Dale, W. Liang, D. Eliyahu, V. Ilchenko, A. B. Matsko, L. Maleki, and C. W. Wong, "Chasing the thermodynamical noise limit in whispering-gallery-mode resonators for ultrastable laser frequency stabilization," arxiv.org/abs/1701.05285
15. I. S. Grudinin, A. B. Matsko, A. A. Savchenkov, D. Strekalov, V. S. Ilchenko, and L. Maleki, "Ultrahigh Q crystalline microcavities," Opt. Commun. **265**, 33-38 (2006)
16. A. A. Savchenkov, A. B. Matsko, V. S. Ilchenko, N. Yu, and L. Maleki, "Whispering-gallery-mode resonators as frequency references. II. Stabilization," *J. Opt. Soc. Am. B.* **24**, 2988-2997 (2007)
17. G. D. Domenico, S. Schilt, and P. Thomann, "Simple approach to the relation between laser frequency noise and laser line shape," App. Opt. **49**, 4801-4807 (2010)